\documentstyle[12pt,epsfig]{article}
\newcommand{\bi}{\bibitem} 
\newcommand{\Two}{T_{20}}
\begin{document}
\begin{center}
{\bf
SCALING BEHAVIOUR OF TENSOR ANALYSING POWER ($A_{yy}$) IN
THE INELASTIC SCATTERING OF RELATIVISTIC DEUTERONS.
}
\vskip 5mm
P.P. Korovin$^{\dag}$, L.V. Malinina, E.A. Strokovsky
\vskip 5mm
{\small {\it
JINR, 141980, Dubna, Moscow region, Russia
}
\\
$\dag$ {\it
E-mail: korovin@sunhe.jinr.ru
}}
\end{center}
\vskip 5mm
\begin{center}
\begin{minipage}{150mm}
\centerline{\bf Abstract}
{\small We suggest a new dimensionless relativistic invariant variable
${\cal{R}} = \Delta{m}_X/\nu$ which may be
interpreted as the ratio
of the excitation energy to the full transferred energy; therefore this
variable measures a "degree of inelasticity" of the scattering.

Existing data on the tensor analysing power of the
$p(\vec{d},d^{\;\prime})X$
and $^{12}C(\vec{d},d^{\;\prime})X$ inelastic scattering at momenta
from 4.2 to 9 GeV/c are analysed in terms of this variable.

We observe that $A_{yy}$ taken as a function of ${\cal{R}}$
does not depend upon the incident energy, the scattering angle (up to the
angles of $\vartheta_{cm}\sim 30^{\circ}$); there is
no noticeable  difference between the proton and nuclear targets as well.

It is remarkable that $A_{yy}$ is maximal (of $\sim 0.5$) when
${\cal{R}} \sim 0.5 - 0.6$
and is small in absolute value when ${\cal{R}}$ is close to its limiting
values of $0$ and $1$.
}

\end{minipage}
\end{center}
\vskip 10mm

\section{Introduction.}

The set of experiments measuring the tensor analyzing power
($A_{yy}$) of inelastic $(\vec{d},d^{\;\prime})$ scattering at the lab.
angles of $0^{\circ}$ and $\sim 5^{\circ}$ off protons and carbon nuclei
in the deuteron momentum range from 4 to 9 GeV/c was performed in Dubna at
1994 - 97 (refs.\cite{data1,data2,data3}).  It was stressed in
refs.\cite{data1,data2,yafmeson} that the region of initial deuteron
momentum of $\sim$ 3 to 9 GeV/c is the optimal one for studies of the
lowest baryon resonances such as $\Delta(1232)$ and $N^*(1440)$. Data on
polarization characteristics of these reactions are of a special interest
because of the "spin-isospin filtering" (see ref.\cite{yafmeson} and
references therein) of different mechanisms, what can be used for better
understanding of the mechanisms of the resonance excitations and
properties of the relevant resonances.

The $A_{yy}$ data published in refs.\cite{data1,data2,data3} were obtained
at different energies and angles; therefore $t$, the 4-momentum transfer
squared, was used in order to analyse and compare data obtained at different
kinematical conditions. It was noticed that $A_{yy}$ plotted versus $t$
demonstrates an approximate scaling (see Fig.1 and refs.\cite{data1,data2}).
It means that at different momenta of initial deuterons the behaviour of
the $A_{yy}(t)$ is approximately the same. At the lab. scattering angle
of $0^{\circ}$ the tensor analyzing power $T_{20} = -\sqrt{2}\cdot A_{yy}(t)$
is negative in the explored $t$-interval ($0 < -t < 0.6~GeV^2/c^2$).
It is small in absolute value at small $-t$ and at $-t > 0.4 GeV^2/c^2$;
the absolute value reaches it's maximum at $0.2 < -t < 0.4~GeV^2/c^2$.
Moreover the approximately universal behaviour of $T_{20}(t)$, or the
scaling, was observed not only at different momenta of initial deuterons
but when the deuteron is scattered on proton or carbon targets.

\begin{figure}[h]
\centerline{\protect\hbox{\epsfig{file=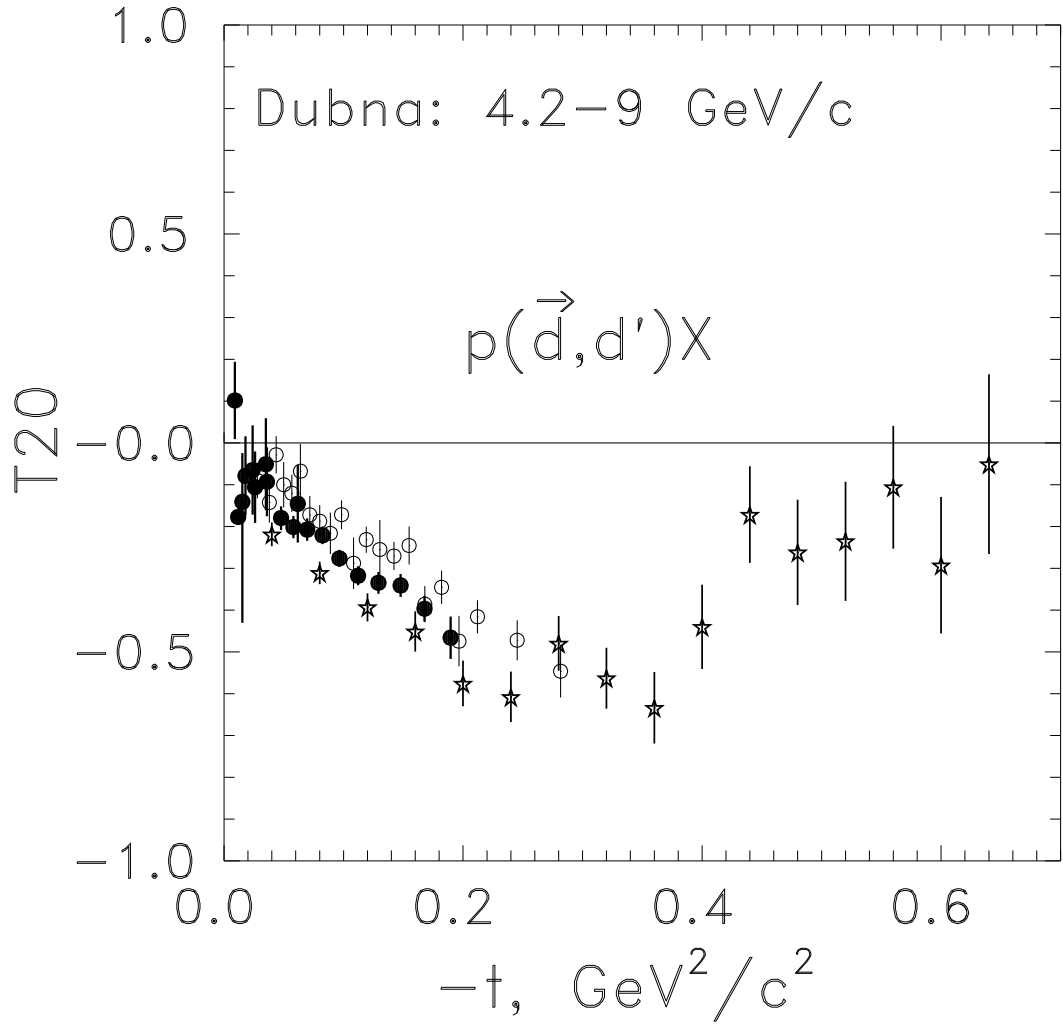,
width=14.00cm}}}
\label{fig1}
\caption[ ]{ $T_{20}(t)$ for $p(\vec{d},d^{\prime})X$ from
ref.\cite{data2}.  Open circles: 4.2-4.5 GeV/c; full circles: 5.53 GeV/c;
stars: 9 GeV/c.}
\end{figure}

Still, it was noticed that the scaling is not perfect: data sets taken at
different energies but at the same scattering angle are systematically
shifted as a whole relative to each other (see Fig.\ref{fig1}). This shift
is small and comparable with the error bars, but noticeable. On the other
hand, existence of the approximate scaling pointed on a possibility that a
better scaling relative to a better variable might be found.

In this paper we suggest such variable. It is relativistic
invariant, dimensionless and has rather clear interpretation; these
features make this variable rather attractive.

Let us define the relativistic invariant dimensionless quantity
\begin{equation}
{\cal R} = \frac{\Delta{m}_X}{\nu},
~~~~~\nu = \frac{1}{m_t}{\cal{P}}_t({\cal{P}}_{d}-{\cal{P}}_{d^\prime})
= m_d u_t(u_d - u_{d^\prime}),
\label{Rdef0}\end{equation}
where ${\cal{P}}_{d}$, ${\cal{P}}_{d^\prime}$ and ${\cal{P}}_t$
are 4 - momenta of the projectile, the ejectile and the target respectively;
$u_d$, $u_{d^\prime}$ and $u_t$ are the 4 - velocities of these particles.
The $\Delta{m}_X = m_X - m_t$  is the difference between masses of the
recoiled system in the final state (the missing mass, $m_X$) and the
initial state (the target mass, $m_t$) respectively.  In other words, this
difference is the energy absorbed by internal degrees of freedom of the
colliding particles (obviously, for elastic scattering one has
$\Delta{m}_X = 0$, hence ${\cal{R}}=0$ in this special case).

It is interesting that ${\cal{R}}$ can be rewritten in the form which
reminds variables widely used in analysis of lepton
deep inelastic scattering:  $$m_X^2 = m_t^2 + t + 2m_t\nu \  \  {;} \  \
{\cal{R}} +
\frac{\Delta{m}_X^2 - t}{2m_t\nu} = 1 ~~~~~\mbox{or}~~~ {\cal{R}}(1 +
\frac{\Delta{m}_X}{2m_t}) = 1 + \frac{t}{2m_t\nu}$$

\begin{figure}[h]
\centerline{\protect\hbox{\epsfig{file=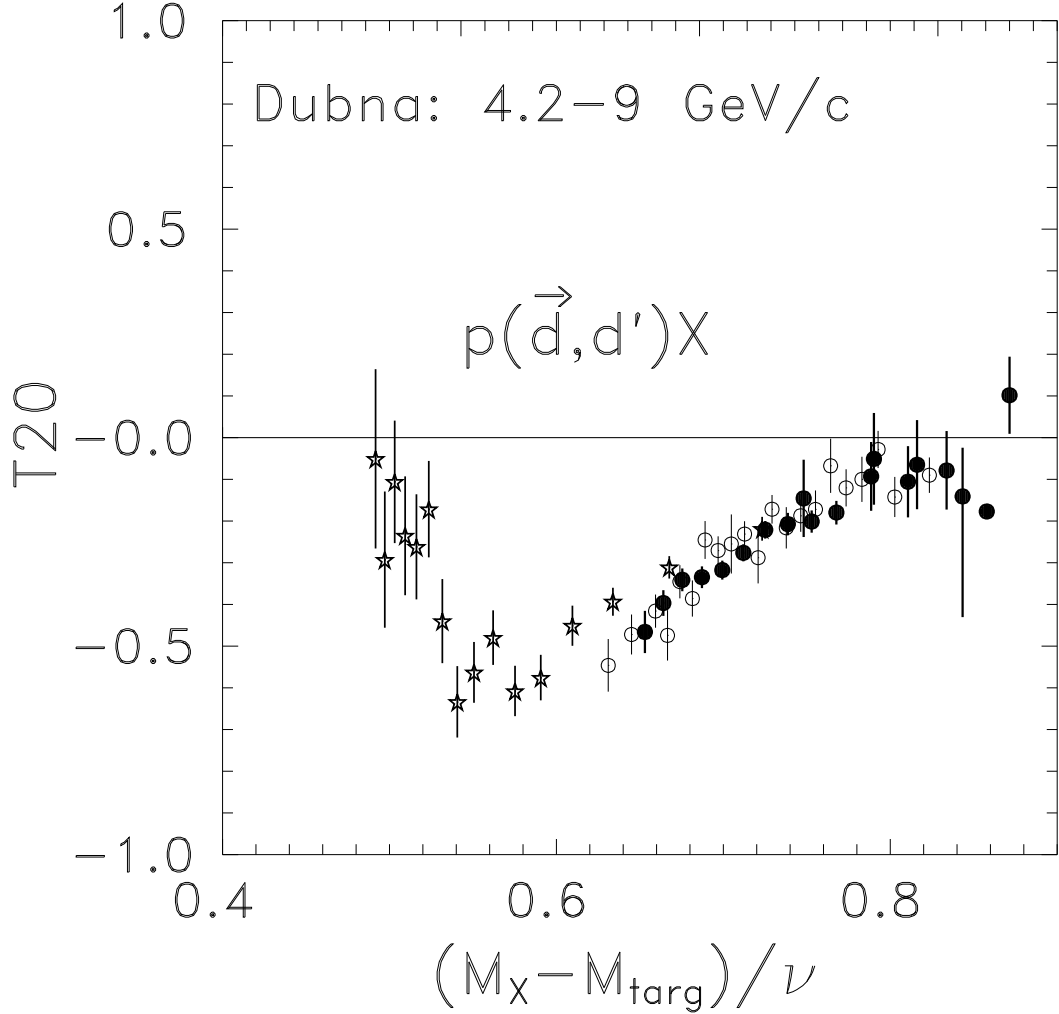,
width=14.00cm}}}
\label{fig2}
\caption[ ]{Data from the Fig.1 plotted versus ${\cal{R}}$. }
\end{figure}

It is easy to see that {\it in the target rest frame}
$${\cal{R}} = \frac{\Delta{m}_X}{Q} = 1 - \frac{T_X}{Q},$$
(where $Q$ is the energy transfer from the projectile to the target
and $T_X$ is kinetic energy of the recoiled system).

Therefore it is possible to interpret ${\cal R}$ as the part of transferred
energy which was absorbed by constituents of the target system. In other
words, this variable can be considered as a measure of the {\it inelasticity}
of the scattering: it differs from zero {\bf only for an inelastic scattering}.


\section{Behaviour of the analysing power as a function of the
inelasticity variable ${\cal R}$.}

First, it is necessary to emphasize that no assumptions about reaction mechanisms,
structure of fragments and so on are made in this paper.

The tensor analyzing power $T_{j\mu}$ can be defined for reactions where
the incident particle has spin larger than $1/2$ as follows:
\begin{eqnarray}
T_{j\mu} =
~Tr\{MS_{j\mu}M^+\}/~Tr\{MM^+\},
\nonumber\end{eqnarray}
where the $M$ is the scattering amplitude. For the case of spin $1$ it can
be written in terms of the reaction cross sections $\sigma_+$, $\sigma_-$ and
$\sigma_0$ for states with
the spin projections onto the quantization axis $S_z = +1 \  {,} 0 \  {,} -1$
respectively as follows:
$$ \Two =\frac{1}{\sqrt{2}}\cdot \frac{\sigma_+ + \sigma_- - 2\cdot
\sigma_0}{\Sigma} \  \  {;} \  \
\Sigma = \sigma_+ + \sigma_- +  \sigma_0$$

The cross section can be expressed in terms of the spherical tensor operators
according Madison Convention \cite{Madison71}.

The data on $T_{20}$ published in refs.\cite{data1,data2}
for the inelastic $p(\vec{d},d^{\prime})X$ scattering of relativistic
deuterons at $\vartheta_{lab}=0^{\circ}$ are plotted on the
Figure \ref{fig2} versus ${\cal{R}}$.

In contrast with the same data plotted versus $t$ on Figure \ref{fig1},
there is no visible tendency of a systematic shift between
data taken at different incident energies on the Figure \ref{fig2}. All
the data show an universal dependence on ${\cal{R}}$; the absolute value
of $T_{20}$ has a maximum at ${\cal{R}} \sim 0.5 - 0.6$.

Apart from the data taken at $\vartheta_{lab} = 0^{\circ}$ from
refs.\cite{data1,data2}, recently a new set of the data on the analysing
power for inelastic scattering of deuterons off carbon nuclei at 9 GeV/c were
published in ref.\cite{data3}. These data were taken at $\vartheta_{lab}
\sim$ 85 mrad (i.e. $\vartheta_{cm} \sim$ 27$^{\circ}$ - 35$^{\circ}$).

Because at the scattering angles larger than $\vartheta_{lab} = 0^{\circ}$
not only $T_{20}$ enters in the cross sections for polarized particles of
spin $1$, it is more convenient to use so-called "carthesian"
representation of the analysing powers. In the experiment of
ref.\cite{data3} the analysing power $A_{yy}$ was actually measured.
Fortunately, at $\vartheta_{lab} = 0^{\circ}$ the $A_{yy}$ is related with
$T_{20}$ in a rather simple way:  $$A_{yy} = - \frac{1}{\sqrt{2}}T_{20}$$
what makes it possible to plot all the available data versus ${\cal{R}}$
on Figure 3. Calculating ${\cal{R}}$, we assume quasifree $d+p$
kinematics, i.e $m_t$ in eq.~(1) is the nucleon mass
as was in the case of $p(d,d{'})X$ scattering.

\begin{figure}[ht]
\centerline{\protect\hbox{\epsfig{file=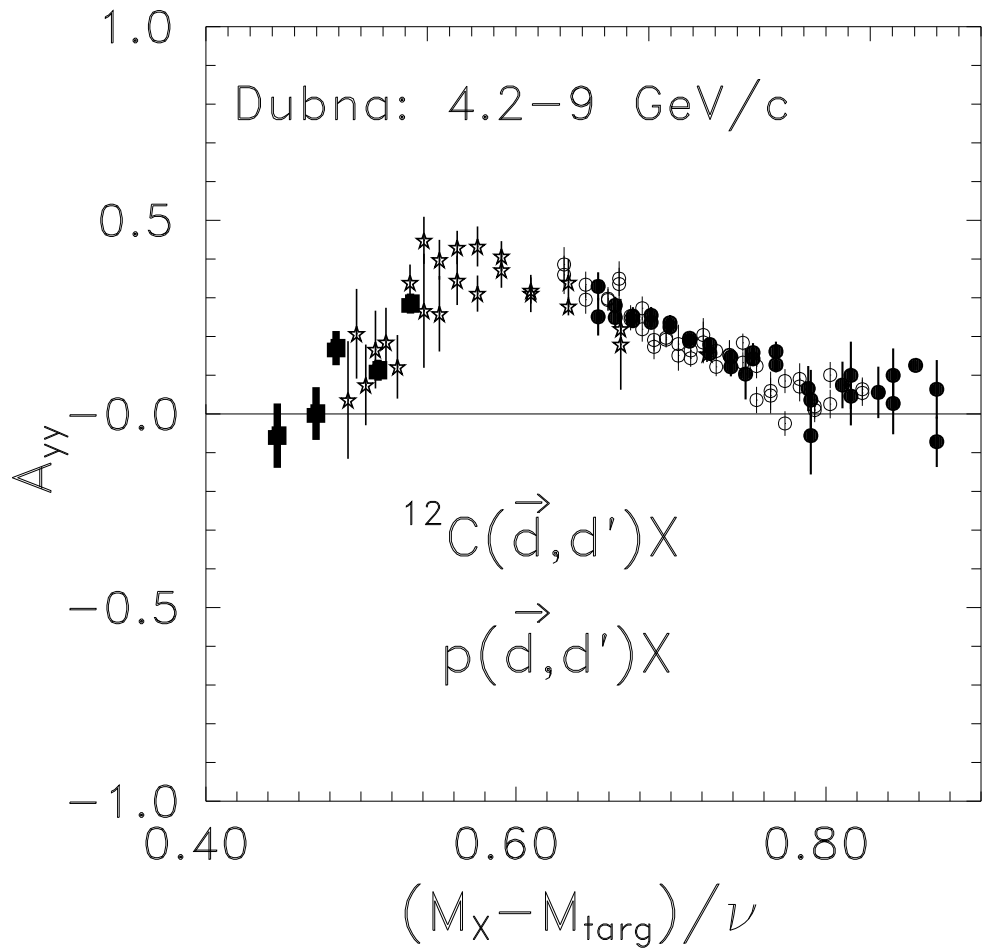,
width=13.00cm}}}
\label{fig3}
\caption[ ]{ $A_{yy}({\cal{R}})$ for
$p(\vec{d},d^{\prime})X$ and $^{12}C(\vec{d},d^{\prime})X$
inelastic scattering; data are taken from refs.\cite{data2,data3}.
Open circles: 4.2-4.5 GeV/c; full circles: 5.53 GeV/c;
stars: 9 GeV/c; full squares: 9 GeV/c at 85 mrad scattering angle; carbon
target\cite{data3}.}
\end{figure}

As one can see from the Figure 3, the data taken at 85 mrad support
the general tendency in the behaviour of tensor analysing power,
which was observed at zero angle.

\section{Conclusion.}


We have suggested a new relativistic invariant and dimensionless variable
for inelastic processes, which takes a constant value equal to zero for any
elastic scattering process.
This variable may be interpreted as a ratio of the excitation energy
to the full transferred energy taken in the target rest frame. Therefore
it is a measure of the degree of "inelasticity" of the scattering process;
in this aspect it reminds the similar parameter introduced in the
ref.\cite{ggg}.

We see that $A_{yy}$ taken as function of $\cal{R}$ does not dependent upon the
initial momentum, the scattering angle and sort of the target.
We see also that when the transferred energy is shared in almost equal
proportions between the internal degrees of freedom of the collidind
particles and the kinetic motion of the recoiled system as whole, the
$A_{yy}$ is maximal.

{\bf This observation inspires an assumption} that this might be a
general feature of the inelastic reactions with polarized particles:  when
the ratio between the "absorbed" and transferred energies is close to 0.5
- 0.6, the polarization effects are strong, while when this ratio is close
to its limits (0 and 1), the polarization effects are weak.
\vspace{0.4cm}

{\it Acknowledgments.} The authors are gratefull to F.A. Gareev
for interest and fruitfull discussions.
This work was supported in part by INTAS-RFBR grant 95-1345.

\section{Appendix.}


In the case of elastic scattering, the ratio ${\cal{R}}$ is zero identically.
Therefore this is "genuine inelastic" variable.
On the other hand, it is easy to see that when one starts from the inelastic
scattering at $\vartheta_{lab}=0$ and approaches to the elastic limit
$m_X \rightarrow m_t$ keeping the scattering angle fixed, one gets
${\cal{R}}\rightarrow 1$ but crosses the unphysical region where
$m_X - m_t <m_{min}$, i.e. where inelastic processes are kinematically forbidden
because the $m_{min}$ is the mass of the lightest particle which can be
created in the reaction under consideration.
At the same time, ${\cal{R}}\rightarrow 0$
if $m_X = m_t$ (fixed) and $\vartheta_{lab} \rightarrow 0$.

That means that there exist a "singular point", because the value of the limit
depends upon the way of approaching that point.

On the other hand, in the completely inelastic limit (the missing mass is at
its maximal value allowed by the conservation laws at given energy, i.e.
$m_X = \sqrt{S} - m_{d^\prime}$), the lab. scattering angle is zero.
Therefore
\begin{equation}
{\cal{R}} = 1 -
\frac{(\sqrt{S}-m_{d^\prime})(\frac{E}{\sqrt{S}}-1)}
{(\sqrt{S}-m_{d^\prime})\frac{E}{\sqrt{S}}-m_{t} }
\longrightarrow \frac{\sqrt{S}}{E}
~~~~~(\mbox{when } \sqrt{S} \gg m_{d^\prime},~m_{t})
\label{InelLimInf}\end{equation}
in the target rest frame ($E$ is full energy).
The ratio ${\cal{R}}$ goes to zero when initial momentum increases to infinity.
\vspace{3mm}

Except for the completely inelastic limit, at fixed initial energy
and scattering angle (in the target rest frame)
the $\cal{R}$ as function of $m_X$ (or $\Delta{m}_X$) has two branches,
which correspond to the "forward" and "backward"
(in the center of mass frame) scattering.
It is clear that the "forward" value of $\cal{R}$ must be larger
than "backward" one at given $\Delta{m}_X$.


\end{document}